\documentclass[acmtocl]{acmtrans2m}

\newtheorem{theorem}{Theorem}[section]

\newdef{definition}[theorem]{Definition}
\newdef{remark}[theorem]{Remark}

\title{Analysis of the deterministic polynomial time solvability of the \textit{0-1-Knapsack} problem.}
\author{JERRALD MEEK}

\begin{abstract}
Previously the author has demonstrated that a representative polynomial search partition is required to solve a \textit{NP-complete} problem in deterministic polynomial time. It has also been demonstrated that finding such a partition can only be done in deterministic polynomial time if the form of the problem provides a simple method for producing the partition.\par

It is the purpose of this article to demonstrate that no deterministic polynomial time method exists to produce a representative polynomial search partition for the Knapsack problem.
\end{abstract}


\begin{document}

\begin{bottomstuff}
	\begin{flushright}
		Jerrald Meek Copyright \copyright 2008
	\end{flushright}
\end{bottomstuff}

\maketitle

\section{Introduction.}
In the article \textit{P is a proper subset of NP} \protect\cite{meek1} the present author proved that a \textit{NP-complete} problem could only be solved in deterministic polynomial time if a representative polynomial search partition can be found in polynomial time. In that same article it has also been demonstrated that finding such a partition by exhaustion requires exponential time on a deterministic machine.\par

It is then clear that any deterministic polynomial time algorithm for a \textit{NP-complete} problem requires one of the following:
\begin{enumerate}
	\item The existence of a method for finding the solution, or a representative polynomial search partition, in deterministic polynomial time by means of examining the form of the problem.
	\item The existence of a deterministic polynomial time solution for the SAT problem which can translate into a fast solution for all \textit{NP-Complete} problems.
\end{enumerate}
These two options are derived from the fact that they are the only ways to avoid an impossible situation which will be reviewed shortly.\par

The author's previous conclusion that $P \neq$ \textit{NP} was dependant upon the assertion that \textit{NP-complete} problems have no direct means of generating either a solution or a representative polynomial search partition. The grounds for this conjecture was briefly examined, but not proven conclusively. The intent of this work is to demonstrate in detail that no deterministic polynomial time solution to the \textit{0-1-Knapsack} problem exists, if this is representative of all \textit{NP-complete} problems, then $P$ is a proper subset of \textit{NP}.

\section{Preliminaries.}
Previously the author has proven the following theorems which will be assumed correct in this article.\par

\newtheorem{thm_opt}{Theorem 4.4 from \textit{P is a proper subset of NP}. \protect\cite{meek1}}[section]
\begin{thm_opt}
\textbf{\emph{P = NP Optimization Theorem.}}\par

The only deterministic optimization of a \textit{NP-complete} problem that could prove \textit{P = NP} would be one that can always solve a \textit{NP-complete} problem by examining no more than a polynomial number of input sets for that problem.
\end{thm_opt}

\newtheorem{thm_part}[thm_opt]{Theorem 5.1 from \textit{P is a proper subset of NP}. \protect\cite{meek1}}
\begin{thm_part}
\textbf{\emph{P = NP Partition Theorem.}}\par

The only deterministic search optimization of a \textit{NP-complete} problem that could prove \textit{P = NP} would be one that can always find a representative polynomial search partition by examining no more than a polynomial number of input sets from the set of all possible input sets.
\end{thm_part}

The definition of the \textit{0-1-Knapsack} problem used in this article will be based off of that used by Horowitz and Sahni \protect\cite{horowitz}.
\noindent \begin{itemize}
	\item Let $S$ be a set of real numbers with no two identical elements.
	\item Let $r$ be the number of elements in $S$.
	\item Let $\delta$ be a set with $r$ elements such that
		\[ {\delta}_i \in \left\{ {0, 1} \right\} \leftarrow 1 \leq i \leq r \]
	\item Let $M$ be a real number.
\end{itemize}

Then
\[ \sum_{i=1}^r {\delta}_i S_i = M \]

Find a variation of $\delta$ that causes the expression to evaluate \textit{true}.

\section{The process of finding a representative polynomial search partition.}

In \textit{P is a proper subset of NP} \protect\cite{meek1} it has been demonstrated that if all subsets of the problem set $S$ which contain a specific element of $S$ can be eliminated, then the set of input sets that must be examined may be reduced by an exponential amount, however this representative search partition will still be exponential in size. It has also been determined that if all sets that contain some combination, i.e. all subsets of $S$ that contain both $S_1$ and $S_2$, are eliminated then this will also produce an exponential-sized representative search partition.

\subsection{Elimination versus direct search.}
The process of finding a representative polynomial search partition by elimination is assumed to be equivalent to a direct search for the partition.\par

\noindent \begin{itemize}
	\item Let $2^r$ represent the cardinality of the set of all possible variations of $\delta$.
	\item Let $r^k$ represent the cardinality of a representative polynomial search partition, where $k$ is a constant exponent.
\end{itemize}

\begin{proof}
If a representative polynomial search partition is to be found, then the number of elements eliminated can be represented by
	\[ 2^r - r^k \]
This can be verified by evaluating
	\[ r^k = 2^r - \left( {2^r - r^k} \right) \]

It is likely to be the case that the easiest way of finding all elements that are not part of a representative polynomial search partition is by first finding the representative polynomial search partition and then identifying all elements of the superset that are not part of the representative polynomial search partition.\par

\begin{itemize}
	\item Let $S$ be the set of all possible inputs for a \textit{NP-Complete} problem.
	\item Let $P \left( S \right) $ be a representative polynomial search partition of $S$.
	\item $\overline{P} \left( S \right) = \neg P \left( S \right) $; all elements of $S$ not in $P \left( S \right)$.
	\item Let $A \left( S \right) $ be an algorithm which identifies $P \left( S \right) $.
	\item Let $\overline{A} \left( S \right) $ be an algorithm which identifies $\overline{P} \left( S \right) $.
\end{itemize}

Then by definition.
\[ A \left( S \right) \Rightarrow P \left( S \right) \]
\[ \overline{A} \left( S \right) \Rightarrow \overline{P} \left( S \right) \]

Is it \textit{true} that
\[ A \left( S \right) \equiv \neg \overline{A} \left( S \right) \]

\begin{table}[h]
	\begin{tabular}{l}
		$A \left( S \right) \equiv \neg \overline{A} \left( S \right) $ \\
		$P \left( S \right) \equiv \neg \overline{P} \left( S \right) $ \\
		$P \left( S \right) \equiv \neg \left( {\neg P \left( S \right) } \right) $ \\
		\hline
		$P \left( S \right) \equiv P \left( S \right) $ \\
	\end{tabular}
\end{table}

It is then the case that if $A \left( S \right) $ is found, then $\overline{A} \left( S \right) $ can be found by the relation that $  A \left( S \right) = \neg \overline{A} \left( S \right) $.  Also, if $\overline{A} \left( S \right) $ is found, then $A \left( S \right) $ can be found by the same relation.  It then follows that if it is provable that either $A \left( S \right) $ or $\overline{A} \left( S \right) $ does not exist, then both $A \left( S \right) $ and $\overline{A} \left( S \right) $ do not exist.
\end{proof}

\subsection{The difficulty of ensuring that a polynomial search partition is representative.}
In \textit{P is a proper subset of NP} \protect\cite{meek1} the requirements of a representative polynomial search partition are described. A polynomial search partition will only be useful for solving a problem if the search partition is representative of the problem. That is, a representative polynomial search partition has the qualities:
\noindent \begin{longenum}
	\item It must be polynomial in size to allow a polynomial time solution without violating the \textit{P = NP Optimization Theorem}.
	\item It must contain at least one input set that will result in the problem evaluating \textit{true} if such a set exists.
	\item If a representative polynomial search partition is to be used in an algorithm that solves a \textit{NP-complete} problem in deterministic polynomial time, then it must be possible to find the representative polynomial search partition in polynomial time on a Deterministic Turing Machine.
\end{longenum}

If the first requirement were the only qualification of a representative polynomial search partition, then there would be no difficulty in proving \textit{P = NP}. It is the second and third requirements that create the true complexity of the problem.\par

Notice that the union of requirement 1 and requirement 2 implies that the representative polynomial search partition not only must contain at least one input that results in a \textit{true} evaluation, but the algorithm for finding the partition must not require that all input sets that result in a \textit{true} evaluation be a part of the search partition. This is due to the fact that sometimes the number of input sets that evaluate \textit{true} may exceed the size of the representative polynomial search partition.

\subsection{Identifying a representative polynomial search partition in deterministic polynomial time.}
The \textit{P = NP Search Partition Theorem} \protect\cite{meek1} stipulates that a representative polynomial search partition can not be found in deterministic polynomial time by examining all elements of the set of all possible input sets. Doing so would require an exponential amount of time on a Deterministic Turing Machine.\par

There is then a conundrum created by the fact that a deterministic polynomial time algorithm for finding a representative polynomial search partition requires that a representative polynomial search partition must first be found.

\begin{proof}
Assume:
\begin{itemize}
	\item Let S be the set of all possible inputs for a SAT problem.
	\item Let O(x) be a search algorithm which identifies an input set for a SAT problem, from the set x of possible input sets for the problem, which results in that problem being a tautology.
	\item Let P(x) be a search algorithm which identifies a representative polynomial search partition for a SAT problem, from the set x of possible input sets for the problem.
\end{itemize}

\vspace{2ex}
\noindent Then
\begin{longitem}
	\item $\left| S \right| $ is exponential.
	\item $O \left( S \right) $ must run in exponential time as required by the \textit{P = NP Optimization Theorem}.
	\item $O \left( x \right) $ may run in polynomial time if the set $x$ is polynomial in size.
	\item $P \left( S \right) $ must run in exponential time as required by the \textit{P = NP Search Partition Theorem}.
	\item $P \left( x \right) $ may run in polynomial time if the set $x$ is polynomial in size.
\end{longitem}

\vspace{2ex}
\noindent Therefore
\begin{longitem}
	\item $O \left( {P \left( S \right) } \right) \Rightarrow O \left( x \right) $ runs in polynomial time, and $P \left( S \right) $ runs in exponential time.  Therefore $O \left( {P \left( S \right) } \right) $ runs in exponential time.
	\item $O \left( {P \left( {P \left( S \right) } \right) } \right) \Rightarrow O \left( x \right) $ runs in polynomial time, and $P \left( x \right) $ runs in polynomial time, and $P \left( S \right) $ runs in exponential time.  Therefore $O \left( {P \left( {P \left( S \right) } \right) } \right) $ runs in exponential time.
\end{longitem}

As can be seen, in order to search for a representative polynomial search partition in polynomial time, there must already exist a representative polynomial search partition.
\end{proof}

Because this situation can never be satisfied, it follows that the problem of the \textit{P = NP Search Partition Theorem} must be avoided entirely by any algorithm that solves a \textit{NP-complete} problem in deterministic polynomial time.\par

It should then be clear that any deterministic polynomial time algorithm for a \textit{NP-complete} problem must produce a representative polynomial search partition in deterministic polynomial time by generating the elements of the partition from an examination of the original problem. If there is no way to directly generate a representative polynomial search partition from the problem in deterministic polynomial time, then there is no deterministic polynomial time solution to the problem; unless SAT is found to have a deterministic polynomial time solution.

\section{Possible methods for avoiding the \textit{P = NP Partition Theorem} trap.}
Prior to examining any possibilities of a polynomial time algorithm for the \textit{0-1-Knapsack} problem, issues related to avoiding the conundrum of the \textit{P = NP Partition Theorem} will first be examined.\par

The \textit{P = NP Partition Theorem} removes any hope that there exists some search trick that allows a deterministic polynomial time solution for any \textit{NP-complete} problem. It is clear that if any method does exist for a deterministic polynomial time solution for the \textit{0-1-Knapsack} problem, then it must exist within the relationship between the elements of $S$ and the value of $M$; or by finding a solution to SAT.

Fortunately there are a finite number of relationships between the set $S$ and the value $M$.
\noindent \begin{enumerate}
	\item A predictable relationship may exist between the elements of $S$ which may be compared to $M$.
	\item $M$ may possess some quality that can be compared against $S$. 
	\item The elements of $S$ may possess some quality that can be compared against $M$.
\end{enumerate}

\section{Predictable relations between the elements of $S$.}
In this section the relations between the elements of the set $S$ and their effect on the solvability of the \textit{0-1-Knapsack} problem will be examined.

\subsection{A \textit{0-1-Knapsack} problem with a predictable relationship between the elements of $S$.}
\noindent \begin{itemize}
	\item Let $S = \left\{ {1, 2, 4, 8, 16, 32, 64} \right\}$
	\item Let $M = 103$
\end{itemize}

\begin{proof}
For this problem, the solution is no more difficult than converting a base 10 number to a base 2 number.

\noindent \begin{longitem}
	\item 64 is less than 103, therefore 64 is a member of the solution set.
	\item 64 + 32 = 96 which is less than 103, therefore 32 is a member of the solution set.
	\item 96 + 16 = 112 which is greater than 103, therefore 16 is not a member of the solution set.
	\item 96 + 8 = 104 which is greater than 103, therefore 8 is not a member of the solution set.
	\item 96 + 4 = 100 which is less than 103, therefore 4 is a member of the solution set.
	\item 100 + 2 = 102 which is less than 103, therefore 2 is a member of the solution set.
	\item 102 + 1 = 103 which is equal to 103, therefore 1 is a member of the solution set.
\end{longitem}

A solution set was found that equals $M$.
\[ 1 + 2 + 4 + 32 + 64 = 103 \]

\vspace{1ex}
Given this example, it should be acceptable without further proof that the \textit{0-1-Knapsack} problem may have a polynomial time solution when there is a predictable relation between the elements of $S$.
\end{proof}

In article 4 \protect\cite{meek4}, the preceding problem will be examined further.  Including a formal proof that base conversion is \textit{NP-Complete}.

\subsection{A \textit{0-1-Knapsack} problem without a predictable relationship between the elements of $S$.}
Although it may be possible to solve the \textit{0-1-Knapsack} problem in deterministic polynomial time when there is a known relation between the elements of $S$, this will not always hold when the elements of $S$ have a random relation.
\begin{itemize}
	\item Let $S$ be a set of randomly selected real numbers.
	\item Let $M$ be a real number.
\end{itemize}

\begin{proof}
If for example $S = \left\{ {3, 10, 14, 21, 23, 26, 32} \right\}$ and $M = 103$. Then the previous algorithm would run as follows.
\noindent \begin{longitem}
	\item 32 is less than 103, therefore 32 is a member of the solution set.
	\item 32 + 26 = 58 which is less than 103, therefore 26 is a member of the solution set.
	\item 58 + 23 = 81 which is less than 103, therefore 23 is a member of the solution set.
	\item 81 + 21 = 102 which is less than 103, therefore 21 is a member of the solution set.
	\item 102 + 14 = 116 which is greater than 103, therefore 14 is not a member of the solution set.
	\item 102 + 10 = 112 which is greater than 103, therefore 10 is not a member of the solution set.
	\item 102 + 3 = 105 which is greater than 103, therefore 3 is not a member of the solution set.
\end{longitem}

No solution set was found, therefore none exists.\par

\vspace{2ex}
The conclusion from this algorithm is incorrect.
\[ 3 + 10 + 14 + 21 + 23 + 32 = 103 \]

\vspace{1ex}
It has been demonstrated that a deterministic polynomial time algorithm may exist for the \textit{0-1-Knapsack} problem when the elements of $S$ are restricted to always possess some predictable relation. However, if the relation between the elements of $S$ is random, then by the very nature of a random relation it is not predictable. It is then the case that any algorithm that relies on a predictable relation between the elements of $S$ will not work for all instances of the Knapsack problem.
\end{proof}

\begin{theorem}
\textbf{\emph{Knapsack Random Set Theorem.}}\par

Deterministic Turing Machines cannot exploit a random relation between the elements of $S$ to produce a polynomial time solution to the \textit{Knapsack} problem.
\end{theorem}

\section{Qualities of $M$.}
In this section qualities of $M$ and their effect on the solvability of the \textit{0-1-Knapsack} problem will be examined.

\subsection{A quality of $M$ that can be compared to $S$.}
If a set of numbers sums to some number $M$, then that set is said to be a composition of $M$. Essentially the \textit{0-1-Knapsack} problem is the problem of finding a composition of $M$ within the set $S$. The most obvious quality of $M$ that can be compared to $S$ is the compositions of $M$.\par

If $M$ and the elements of $S$ can be any real number, then $S$ must be compared to the real compositions of $M$. The set of real compositions of $M$ is infinite.
\begin{proof}
For example, if $M = 5$ then:\newline
\begin{center}
	5 = 1 + 4, or 5 = 1.1 + 3.9, or 5 = 1.2 + 3.8
\end{center}
There are an infinite number of real numbers between 1 and 2 and an infinite number of real numbers between $M - 2$ and $M - 1$. Therefore, the number of real compositions of $M$ which contain only two elements such that each element exists in the ranges 1 to 2 or $M - 2$ to $M - 1$ is infinite. It is therefore easy to see that the number of real compositions of $M$ is also infinite; compairing them to the elements of $S$ would require infinite time.
\end{proof}

If $M$ and the elements of $S$ are restricted to integers, then $S$ must be compared to the integer compositions of $M$. The set of integer compositions of $M$ is infinite.
\begin{proof}
Again let $M = 5$, then:
\begin{center}
	5 = 6 + (-1), or 5 = 7 + (-2), or 5 = 8 + (-3)
\end{center}
There are an infinite number of negative integers. For each negative integer there exists a positive integer such that the sum of the negative and the positive integer equals $M$. Therefore, the number of integer compositions of $M$ which contain only two elements is infinite. It is then easy to see that the number of integer compositions of $M$ is also infinite; compairing them to the elements of $S$ would require infinite time.
\end{proof}

If $M$ and the elements of $S$ are further restricted to whole numbers, then $S$ must be compared to the whole number compositions of $M$, which are finite. With the absence of negative numbers, any element of $S$ greater than $M$ will not be an element of any whole number composition of $M$.\par

The problem of determining if a set is a subset of another set can be solved in deterministic polynomial time. Therefore, when $M$ and the elements of $S$ are restricted to whole numbers and all whole number compositions of $M$ are known, then the \textit{0-1-Knapsack} problem could be solved in deterministic polynomial time.\par

This method will work only under two conditions. One being that the number of compositions of $M$ must be polynomial relative to the cardinality of $S$ and the other is that the whole number compositions of $M$ can be found in deterministic polynomial time. However, both of these conditions are false. The number of whole compositions of $M$ can be represented as $2^{M - 1}$, and grows exponentially as $M$ increases. Also, the problem of finding all compositions of $M$ is the same thing as finding all input sets that evaluate \textit{true} for a \textit{0-1-Knapsack} problem where $S$ contains all whole numbers less than or equal to $M$.\par

\begin{proof}
Notice that if $M$ is a small number, for example 5, then
\[ \Sigma = \left\{ {1, 1, 1, 1, 1, 2, 2, 3, 4, 5} \right\} \]

\noindent the problem of finding the compositions of $M$ is the problem of finding all variations of $\Delta$ that result in a \textit{true} evaluation of the problem
	\[ \sum_{i=1}^{2^{\left| \Sigma \right| }} \Delta_i \Sigma_i = 5 \]

\noindent Computations Required: 10($2^{10}$) = 10,240\newline
\noindent Compositions Found: $2^{5 - 1}$ = 16\par

The process of finding all solution sets to this problem and determining if any are subsets of $S$ may be faster than solving the problem by standard means when there are several elements of $S$, and $M$ is small.
\end{proof}

Notice that this optimization may not work when $M$ is large.  If $M = 45,182$, and the set $S$ only contains 15 elements, then obviously the original problem will be easier to solve than the problem of finding the compositions of M.\par

It has been demonstrated that examining compositions of $M$ can only work when $M$ and the elements of $S$ are restricted to whole numbers. Furthermore, this method does not always produce a faster means of solving the \textit{0-1-Knapsack} problem.

\begin{theorem}
\textbf{\emph{Knapsack Composition Theorem.}}\par

Compositions of $M$ cannot be relied upon to always produce a deterministic polynomial time solution to the \textit{0-1-Knapsack} problem.
\end{theorem}

\subsection{Other qualities of $M$ that can be compared to $S$.}
It is possible for other qualities of $M$ to be found that could be compared to $S$. However, $S$ is a set of numbers that represent elements of possible compositions of $M$. Also, the subset that is being looked for is a composition of $M$. It is then the case that any quality of $M$ that can be compared to $S$ must ultimately produce a composition of $M$.\par

Furthermore, qualities of $M$ are not related to $S$. If $M$ remains constant and $S$ changes then $M$ retains all qualities previously held by $M$. It is then obvious that if a quality of $M$ can be used to find a composition of $M$ that is in $S$, then that quality must be capable of generating any composition of $M$.\par

\begin{theorem}
\textbf{\emph{Knapsack $M$ Quality Reduction Theorem.}}\par

Any quality of $M$ that could be used to find a composition of $M$ within $S$ would be equivalent to finding all compositions of $M$.
\end{theorem}

\section{Qualities of the elements of $S$ that can be compared to $M$.}
The most obvious quality of the elements of $S$ that can be compared to $M$ is that $S$ may or may not contain a composition of $M$. In fact, determining this quality is exactly what the \textit{0-1-Knapsack} problem is.\par

Another quality of the elements of $S$ could exist within a predictable relation between the elements of $S$; however this relation has already been examined.\par

Any quality of the elements of $S$ not equivalent to the previous two suffers form the problem that if $S$ remains constant and $M$ is changed, then the quality of $S$ will remain unchanged and may no longer apply to $M$. It is then the case that any quality of the elements of $S$ that identifies a representative polynomial search partition must be applicable to all possible values of $M$. The existence of such a quality is absurd.
\begin{itemize}
	\item Let $S = \left\{ {1, 16, 43, 102} \right\} $
\end{itemize}

\begin{proof}
All subsets of $S$ are\newline
\vspace{1ex}

\begin{acmtable}{322pt}[h]
	\centering
	\begin{tabular}{|l|l|l|l|l|l|}
		$ \oslash $ & $ \left\{ 1 \right\} $ & $ \left\{ 16 \right\} $ & $ \left\{ 43 \right\} $ & $ \left\{ 102 \right\} $ & \\
		\hline
		$ \left\{ {1, 16} \right\} $ & $ \left\{ {1, 43} \right\} $ & $ \left\{ {1, 102} \right\} $ & $ \left\{ {16, 43} \right\} $ & $ \left\{ {16, 102} \right\} $ & $ \left\{ {43, 102} \right\} $ \\
		\hline
		$ \left\{ {1, 16, 43} \right\} $ & $ \left\{ {1, 16, 102} \right\} $ & $ \left\{ {1, 43, 102} \right\} $ & $ \left\{ {16, 43, 102} \right\} $ & & \\
		\hline
		$ \left\{ {1, 16, 43, 102} \right\} $ & & & & & \\
	\end{tabular}
	\caption{Subsets of $S$}
\end{acmtable}

\vspace{1ex}

The summations of these sets are\newline
\vspace{1ex}

\begin{acmtable}{148pt}[h]
	\centering
	\begin{tabular}{|l|l|l|l|l|l|}
		0 & 1 & 16 & 43 & 102 & \\
		\hline
		17 & 44 & 103 & 59 & 118 & 145 \\
		\hline
		45 & 119 & 146 & 161 & & \\
		\hline
		162 & & & & & \\
	\end{tabular}
	\caption{Sums of subsets of $S$}
\end{acmtable}

Notice that in this example no two subsets of $S$ sum to the same number. It is then the case that a quality of $S$ must be able to produce any one of an exponential number of possible values for $M$. This results in a search problem over an exponential set.
\end{proof}

\begin{theorem}
\textbf{\emph{Knapsack Set Quality Theorem.}}\par

Using any quality of the elements of $S$ to solve the \textit{0-1-Knapsack} problem will be no less complex than the standard means of solving the \textit{0-1-Knapsack} problem.
\end{theorem}

\section{The Backpacker's Card Game.}
Up to this point, various methods have been examined which are representative of all possible methods for optimizing the \textit{0-1-Knapsack} problem from relations between the set $S$ and the value of $M$. Some of these methods do produce optimizations under limited conditions, but none produce optimizations under all conditions.\par

Because special conditions exist where the \textit{0-1-Knapsack} problem can be solved in deterministic polynomial time; it is necessary to prove that a problem exists where none of these special conditions are present.\par

In this section we examine the results of a polynomial algorithm for a FNP problem. For this analysis, a problem will be introduced which is related to the \textit{0-1-Knapsack} problem.\par

The \textit{Backpacker's Card Game} problem involves one or more decks of cards with 52 cards in a deck, and two 6 sided dice. The cards will each be assigned a numeric value derived from rolling the dice. The first card will have a value from 2 to 12, the second card will have a value equal to that of the first card plus a random value from 2 to 12. This method will be continued until all cards have an assigned value.
\begin{itemize}
	\item Let $S$ be the set of all numeric values represented by the cards in the deck.
	\item Let $r$ be the number of elements in $S$.
	\item Let $\delta$ be a set with $r$ elements such that
		\[ \delta \in \left\{ {0, 1} \right\} \leftarrow 1 \leq i \leq r \]
	\item Variations of $\delta$ are restricted to those having exactly 5 elements equal to 1.
	\item Let $M$ be the sum of the values of 5 cards that are drawn from the deck at random.
\end{itemize}

\begin{proof}
Given the value of $M$ and the values of all cards, find in deterministic polynomial time any set of cards that sum to $M$. The problem can then be represented by the expression
	\[ \sum_{i=1}^r {\delta}_i S_i = M \]
Find one variation of $\delta$ that causes this expression to evaluate \textit{true}.\par

The number of possible 5 card combinations from a 52 card deck is represented by $\left( \stackrel{52}{5} \right) = \frac{52!}{\left( {52-5} \right) ! 5!} = 2,598,960$. Notice that if the restriction of only using 5 card hands were not in place then the total number of possible variations of $\delta$ would be $2^{52} = 4,503,599,627,370,496$. It is then obvious that the \textit{Backpacker's Card Game} is a less complex problem than the \textit{0-1-Knapsack} problem.\par

Also, there is no predictable relation between the elements of $S$ that can be depended upon because the elements of $S$ are determined by rolling dice. The lowest possible value of $M$ is 2 + 4 + 6 + 8 + 10 = 30 (when the 5 lowest cards were produced by rolling 2), while the highest possible value is 336 + 348 + 360 + 372 + 384 = 1800 (when all 52 dice rolls produced 12). However, when rolling two six-sided dice the odds of rolling 2 is 1:12, the odds of rolling 12 is 1:12, but the odds of rolling 7 is 1:4. It is then the case that the average lowest value card will be 7 and the average highest value card will be around $7 \times 52 = 364$. The value of $M$ should be expected to most often come somewhere close to $\left( {\frac{52}{2} \times 7} \right) \times 5 = 910$.\par

The problem of finding all compositions of $M$ when $M = 910$ will require more than $2^{52}$ computations.  It will then be easier to solve this problem by standard means.\par

Notice that the \textit{Backpacker's Card Game} is not a \textit{NP-complete} problem. In this problem there is always at least one set that evaluates \textit{true}, but the objective is to find one of these sets. The problem is then a function problem that could be solved in polynomial time on a Non Deterministic Turing Machine. It is then a member of the \textit{FNP} complexity class and is \textit{NP-hard}.
\end{proof}

\subsection{Time requirement for the \textit{Backpacker's Card Game}.}
If the \textit{Backpacker's Card Game} is solvable in deterministic polynomial time, then there exists a function in the form of $r^k$ that represents the maximum number of computations required to solve the problem. By Cook's definition of polynomial time \protect\cite{cook} this should actually be $r^k + k$, however this would make the math much more complex. If we add 1 to $k$ then the second term can be dropped because $r^k > r^{k-1} + \left( {k - 1} \right)$, therefore the maximum number of computations can be bounded by a monomial.\par

At this time it is necessary to recall that a Non Deterministic Turing Machine has the ability to simulate an extremely lucky guesser who always guesses correctly regardless of the odds. This ability is the one difference between deterministic machines and non deterministic machines. Because Deterministic Turing Machines can not guess, it is then the case that any deterministic algorithm must abide by the laws of probability.\par

In the \textit{Backpacker's Card Game} it is guaranteed that there is always one input set that sums to $M$. However, there is no guarantee that there are any additional input sets that sum to $M$. So the worst possible odds that any particular guess will be the set that sums to $M$ happens when there exists only one such set.\par

Therefore, when there is only one set that sums to $M$, the odds of any particular guess being correct can be represented by 1:2,598,960. It is then the case that a deterministic machine will only have a 100\% chance of finding the correct answer after evaluating all 2,598,960 possibilities.\par

\begin{proof}
Let $c$ represent the number of computations required to evaluate a single input set. The deterministic polynomial time algorithm would then require a number of computations that is represented by
	\[ r^k = 2,598,960c \]
For simplicity, $c$ will be assumed equal to 1. Then
	\[ r^k = 2,598,960 \Rightarrow k \approx 3.73822 \]

If this deterministic polynomial time algorithm truly works for the problem, then it must work not only when one deck is used, but also when any numbers of decks are used. If the shuffle is increased to 2 decks then there are 64 cards. In this case the number of possible 5 card hands is $\left( {\stackrel{64}{5}} \right) = \frac{64!}{\left( {64-5} \right) ! 5!} \approx 1.27 \times 10^{84}$.\par

The algorithm uses $64^{3.73822} = 5,648,044$ computations to evaluate $1.27 \times 10^{84}$ possible input sets. This is obviously impossible for a deterministic machine.
\end{proof}

\section{Other possible solutions for the \textit{0-1-Knapsack} problem.}

The possibility for a deterministic polynomial time search algorithm for \textit{NP-complete} problems has been ruled out \protect\cite{meek1}. The possibility of a fast solution from the form of the problem has also been eliminated. There are now only two remaining possibilities. These possible methods will briefly be mentioned here.

\subsection{Algorithms relying on probabilistic methods.}

It may be possible to produce an algorithm that quickly generates input sets that are more likely to be accepted than other input sets. However, the nature of a probabilistic solution is such that sometimes the most likely result will not be correct. This is exactly the reason why the \textit{Rabin-Miller} test requires several runs to determine the primality of a number \protect\cite{hurd}.\par

If a method exists which can generate inputs in descending order of their probability of acceptance then the probability that an accepting input will be found will increase as $n$ increases when $n$ is the number of inputs that have been tried. However, if no accepting input has been found after some polynomial number of attempts, then there is no grantee that an accepting input does not exist within the set that has not been examined.\par

For example, suppose an algorithm can find an accepting input set if such a set exists within a deterministically polynomial bounded time limit with a 99.9\% success rate. Then that means for every 1000 \textit{NP-complete} problems, there is 1 \textit{NP-complete} problem such that this algorithm does not produce a correct solution in deterministic polynomial time.\par

Such a method could have some of the advantages of a \textit{P = NP} algorithm, and would defiantly be useful. However, a probabilistic algorithm would only prove \textit{P = NP} if it can give a 100\% probability of finding an accepting input set within deterministic polynomial time. Such an algorithm would then be equivalent to a solution that produces a representative polynomial search partition.\par

\subsection{A deterministic polynomial time algorithm for SAT.}

Any \textit{NP-complete} problem would be solvable in deterministic polynomial time if SAT had a deterministic polynomial time solution. However, a deterministic polynomial time solution for SAT would only be found by first finding a deterministic polynomial time solution for some other \textit{NP-complete} problem.

\begin{itemize}
	\item Let $a$, $b$, and $c$ be \textit{true} or \textit{false}, but the values are unknown.
	\item $a \Rightarrow x = 1, \neg a \Rightarrow x = 0$.
	\item $b \Rightarrow y = 1, \neg b \Rightarrow y = 0$.
	\item $c \Rightarrow z = 1, \neg c \Rightarrow z = 0$.
\end{itemize}

Notice that the two statements are logically equivalent.
	\[ \left[ {a \vee b \vee c} \right] \equiv \left[ {x + y + z > 0} \right] \]

Therefore, a deterministic polynomial time solution for SAT actually implies that one inequality can produce the value for three unknown variables.\par

Because it is impossible to directly develop a deterministic polynomial time solution to SAT, then the only way to prove that SAT has a deterministic polynomial time solution is by transfering one from another \textit{NP-complete} problem. It is then the case that SAT can only really be proven to have no fast solution by proving that all other \textit{NP-complete} problems have no fast solutions.\par

It is now known that no deterministic polynomial time solution for the \textit{Knapsack} problem exists (unless SAT provides one). It is the purpose of the article \textit{Analysis of the postulates produced by Karp's Theorem.} \protect\cite{meek4} to demonstrate that a deterministic polynomial time solution can not be produced for SAT from another \textit{NP-Complete} problem.  Therefore, $P$ is without a doubt strictly contained within \textit{NP}.

\section{Conclusion.}
It was stated earlier that the \textit{Backpacker's Card Game} is a less complex problem than the \textit{0-1-Knapsack} problem. Assuming SAT has no deterministic polynomial time solution, then because the \textit{Backpacker's Card Game} can not be solved in deterministic polynomial time, it follows that the \textit{0-1-Knapsack} problem is also unsolvable in deterministic polynomial time.\par

\begin{flushright}
Q.E.D.
\end{flushright}

\section{Version History.}
The author wishes to encourage further feedback which may improve, strengthen, or perhaps disprove the content of this article. For that reason the author does not publish the names of any specific people who may have suggested, commented, or criticized the article in such a way that resulted in a revision, unless premission has been granted to do so.\par

\noindent \textbf{arXiv Current Version}\newline
6Sep08 Submitted to arXiv.
\begin{itemize}
	\item Minor revision.
\end{itemize}

\noindent \textbf{arXiv Version 4}\newline
5Sep08 Submitted to arXiv.
\begin{itemize}
	\item Correction to mathmatical errors.
\end{itemize}

\noindent \textbf{arXiv Version 3}\newline
23Aug08 Submitted to arXiv.
\begin{itemize}
	\item Refrance to \protect\cite{meek4} added.
\end{itemize}

\noindent \textbf{arXiv Version 2}\newline
09May08 Submitted to arXiv.
\begin{itemize}
	\item Section 9 was added.
	\item Refrance to \protect\cite{hurd} added.
	\item Minor revisions.
\end{itemize}

\noindent \textbf{arXiv Version 1}\newline
05May08 Submitted to arXiv.

\begin{received}
Received xx/2008; revised xx/2008; accepted xx/2008
\end{received}
\end{document}